# Planck spectroscopy

**Authors:** Yuzhe Xiao[1], Chenghao Wan[1,2], Jad Salman[1], Ian J. Maywar[1], Jonathan King[1], Alireza Shahsafi[1], and Mikhail A. Kats[1,2,3*]

**Affiliations:**

[1]Department of Electrical and Computer Engineering, University of Wisconsin-Madison, Madison, Wisconsin 53706, USA

[2]Department of Materials Science and Engineering, University of Wisconsin-Madison, Madison, Wisconsin 53706, USA

[3]Department of Physics, University of Wisconsin-Madison, Madison, Wisconsin, USA 53706

*Correspondence to: mkats@wisc.edu.

**Abstract:**

All spectrometers rely on some mechanism to achieve spectral selectivity; common examples include gratings, prisms, and interferometers with moving mirrors. We experimentally demonstrated and validated a spectroscopic technique—here dubbed Planck spectroscopy—that measures the spectral emissivity of a surface using only a temperature-controlled stage and a detector, without any wavelength-selective optical components. Planck spectroscopy involves the measurement of temperature-dependent thermally emitted power, where the spectral selectivity is realized via the temperature- and wavelength dependence of Planck's law. We experimentally demonstrated and validated Planck spectroscopy in the mid infrared, for wavelengths from 3 to 13 µm—limited primarily by the bandwidth of our detector—with resolution of approximately 1 µm. The minimalistic setup of Planck spectroscopy can be implemented using infrared cameras to achieve low-cost infrared hyperspectral imaging and imaging ellipsometry.

**Main Text:**

Existing spectroscopy methods rely on some optical component to enable wavelength selectivity, and can be roughly classified into the following three categories (Fig. 1): (A) spectrometers that use a dispersive component, such as a diffraction grating [1], prism [2], or disordered optical medium that features wavelength-dependent scattering [3]; (B) spectrometers that obtain the spectrum by measuring the temporal coherence of light via reconfigurable interferometers, such as Fourier-transform spectrometers (FTSs, a.k.a. FTIRs) [4]; and (C) spectrometers that sample the spectrum using swappable filters such as those in a filter wheel [5], an array of static filters in front of detector pixels [6], [7], a dynamically tunable filter [8], or an array of detectors with different spectral responses [9].

In this report, we investigate a spectroscopic technique—Planck spectroscopy—that requires no wavelength-selective optical component, but instead uses the fundamental temperature- and wavelength dependence of Planck's law of thermal radiation to enable spectroscopic measurement. Though this approach has been proposed previously [10], [11], it to our knowledge has not been validated or quantified in terms of accuracy and resolution. Indeed, our experiments and calculations show that very careful measurements are required to yield meaningful spectra. We experimentally demonstrated Planck spectroscopy over the wavelength range from 3 to 13 µm—limited primarily by the bandwidth of our detector—and validated the resulting measured spectral emissivity curves against measurements with a Fourier-transform spectrometer.

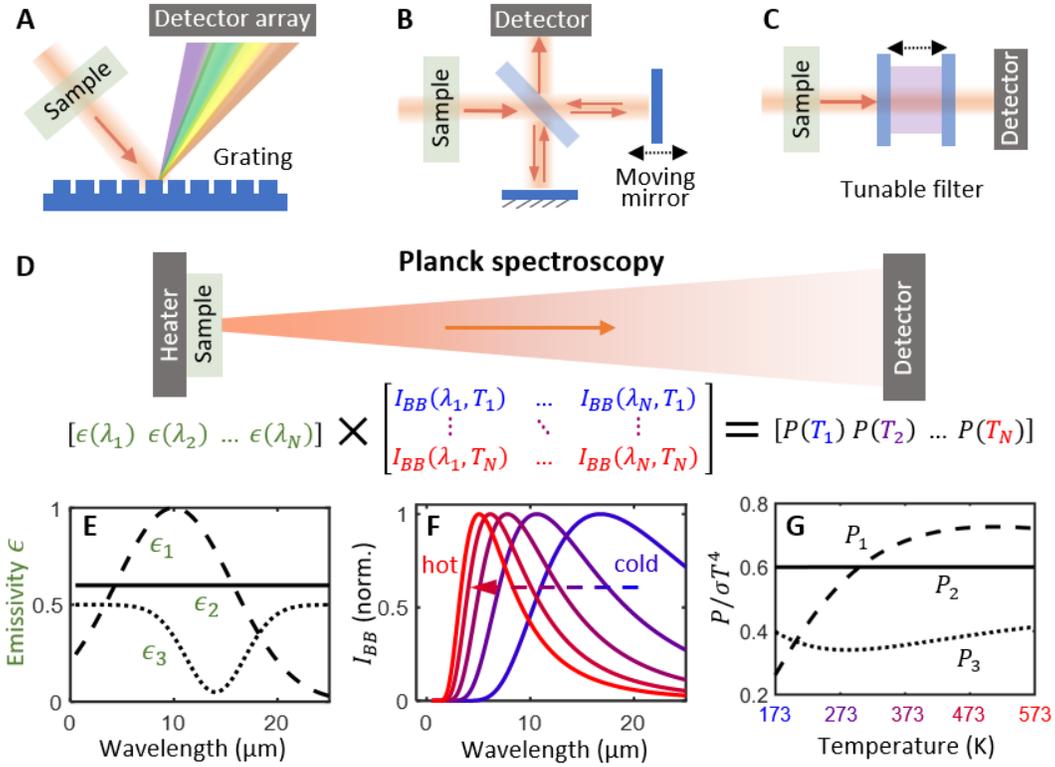

**Fig. 1. Comparison between conventional spectroscopic methods and Planck spectroscopy.** Panels (A-C) depict three common spectroscopy mechanisms using (A) a dispersive component, such as a grating, to spatially separate light with different wavelengths; (B) a moving mirror to modify the interference condition of an interferometer to enable reconstruction of the spectrum; and (C) a tunable filter to select different portions of the spectrum. (D): Planck spectroscopy requires only a temperature stage and a detector. The total emission power $P(T)$ is determined by the sample emissivity $\epsilon(\lambda)$ and the blackbody radiation spectrum $I_{BB}(\lambda, T)$ via a linear relation. $\epsilon(\lambda)$ can be reconstructed computationally from knowledge of $P(T)$ by solving a constrained linear least-squares problem. Due to the temperature-dependent spectral shape of $I_{BB}(\lambda, T)$ (F), objects with different emissivities, $\epsilon(\lambda)$ (E), have different temperature-dependent emitted power, $P(T)$ (G). In (G), the power is normalized by $\sigma T^4$, where $\sigma$ is the Stefan-Boltzmann constant.

To our knowledge, Planck spectroscopy requires fewer optical components than any other spectrometer—consisting at its core of just a temperature controller and a detector (Fig. 1D). The optical properties of the sample are imprinted on the light emitted as thermal radiation, which results from the temperature-dependent stochastic motion of charges that constitute the material [12]. The emitted power spectrum of an object is the product of its thermal emissivity $\epsilon(\lambda)$ and the blackbody radiation spectrum $I_{BB}(\lambda, T)$, which is given by Planck's law [13], [14]. Due to the temperature dependence of the Bose-Einstein distribution [12], the spectral shape of $I_{BB}(\lambda, T)$ is temperature-dependent (Fig. 1F). In particular, the radiance peak shifts to shorter wavelengths as temperature increases, as described by Wien's displacement law [15].

Due to the temperature- and wavelength dependence of $I_{BB}(\lambda, T)$, the emissivity spectrum $\epsilon(\lambda)$ is encoded in the temperature-dependent total emission power, $P(T)$. For example, an object with a constant emissivity over a broad spectral range ($\epsilon_2$, Fig. 1E) has $P(T)$ proportional to $T^4$ ($P_2$, Fig. 1G), as per the Stefan-Boltzmann law [16]. Meanwhile, for objects with wavelength-dependent emissivity ($\epsilon_1$ and $\epsilon_3$, Fig. 1E), $P(T)$ deviates from the Stefan-Boltzmann law, with the shape of $P(T)$ determined by $\epsilon(\lambda)$ ($P_1$ and $P_3$, Fig. 1G). Note that here we assume $\epsilon(\lambda)$ is approximately independent of temperature, which is a reasonable assumption for most solids over temperature ranges of hundreds of degrees, with some notable exceptions [17], [18].

In Planck spectroscopy, varying the temperature changes the spectral shape of $I_{BB}(\lambda, T)$ (Fig. 1E), which is conceptually similar to using a tunable filter (Fig. 1C), except no physical filter is used. Assuming discrete sets of wavelengths, $\lambda_i$—with constant spacing $\Delta\lambda$—and temperatures, $T_j$, the relationship between $P(T)$ and $\epsilon(\lambda)$ can be written as a matrix equation:

$$\begin{bmatrix} P(T_1) \\ \vdots \\ P(T_N) \end{bmatrix} = \begin{bmatrix} I_{BB}(\lambda_1, T_1) & \cdots & I_{BB}(\lambda_N, T_1) \\ \vdots & \ddots & \vdots \\ I_{BB}(\lambda_1, T_N) & \cdots & I_{BB}(\lambda_N, T_N) \end{bmatrix} \begin{bmatrix} \epsilon(\lambda_1) \\ \vdots \\ \epsilon(\lambda_N) \end{bmatrix} \Delta\lambda \tag{1}$$

The $N$ unknown spectral emissivity values, $\epsilon(\lambda_i)$, can be extracted from the measured power $P(T_i)$ at $N$ temperatures by solving a constrained linear least-squares problem, as described below in the context of our experimental demonstration.

Figure 2A depicts the experimental setup we used to demonstrate Planck spectroscopy. Samples were placed on a temperature stage, and the corresponding thermal emission from the sample was collected by a lens from the normal direction and focused into a broadband infrared detector (see more details in Supplementary Materials, Sec. 1). The measured thermal-emission power is represented by the output voltage from the detector, $V(T)$, as

$$V(T) = \int \eta(\lambda)[\epsilon(\lambda)I_{BB}(\lambda, T) + B(\lambda)]d\lambda, \tag{2}$$

where the integration is over the detector bandwidth. Here, $B(\lambda)$ is the background emission from the surrounding environment and $\eta(\lambda)$ is the system response that accounts for the detector responsivity and the collection efficiency along the optical path. In general, background emission has a non-negligible impact on thermal-emission measurements, especially for emitters with low or moderate temperatures [19], [20]. In our experiment, $B(\lambda)$ is largely independent of the sample temperature because the lab room temperature is well maintained, so this term can be eliminated by looking at the differences between measured data at different temperatures and the first temperature, i.e., $\Delta V = V(T_i) - V(T_1)$. Therefore, we reformulated Eq. 2 in terms of voltage differences between temperatures:

$$\begin{bmatrix} \Delta V(T_1) \\ \vdots \\ \Delta V(T_N) \end{bmatrix} = \begin{bmatrix} \Delta I_{BB}(\lambda_1, T_1) & \cdots & \Delta I_{BB}(\lambda_N, T_1) \\ \vdots & \ddots & \vdots \\ \Delta I_{BB}(\lambda_1, T_N) & \cdots & \Delta I_{BB}(\lambda_N, T_N) \end{bmatrix} \begin{bmatrix} \epsilon(\lambda_1) \\ \vdots \\ \epsilon(\lambda_N) \end{bmatrix} \begin{bmatrix} \eta(\lambda_1) \\ \vdots \\ \eta(\lambda_N) \end{bmatrix} \Delta\lambda, \tag{3}$$

where $\Delta V(T_j) = V(T_{j+1}) - V(T_1)$ and $\Delta I_{BB}(\lambda_i, T_j) = I_{BB}(\lambda_i, T_{j+1}) - I_{BB}(\lambda_i, T_1)$ are the differences in measured voltage and the blackbody radiation spectrum, respectively.

The extraction of $\epsilon(\lambda_i)$ requires not only experimentally measured $V(T_j)$, but also knowledge of the system response $\eta(\lambda_i)$, which may not be precisely known. To obtain $\eta(\lambda_i)$, one can solve Eq. 3 using experimental data $V_{ref}(T_j)$ from a reference sample with a known emissivity $\epsilon_{ref}(\lambda_i)$. Then, once $\eta(\lambda_i)$ is determined, Eq. 3 can be used to extract $\epsilon(\lambda_i)$ of an unknown sample.

Furthermore, in the temperature- and wavelength range of interest, Eq. 3 is ill-conditioned because the condition number of the blackbody-radiation matrix is much larger than one [21] (see Supplementary Materials, Sec. 4). Therefore, the solution of Eq. 3 via matrix inversion is not robust against the noise in the measured power. In our case, there are several constraints that can be applied to increase the robustness of the solution: (i) $\epsilon(\lambda)$ must be between 0 and 1, (ii) $\eta(\lambda)$ must be larger than 0, and (iii) both $\epsilon(\lambda)$ and $\eta(\lambda)$ are expected to be smooth functions of wavelength. To solve Eq. 3, we used a linear least-squares solver using (i) and (ii) as constraints and smoothing the solution to satisfy (iii); see more details in Supplementary Materials, Sec. 4. We note that more-sophisticated reconstruction algorithms such as adaptive regularization [9], [22] and principle component regression [23] may be used to better solve Eq. 3.

In Figure 2B, we plotted the normalized $V(T)$, corresponding to the measured thermally emitted power, for five samples, including a laboratory blackbody (a vertically oriented array of carbon nanotubes [24] on a silicon wafer, with constant $\epsilon_{ref} \sim 0.97$ across the mid infrared, calibrated previously in ref. [19]), a sapphire wafer, a fused-silica wafer, and two n-doped silicon wafers with different doping levels. $V(T)$ was measured from 193 to 523 K with a step size of 5 K. We selected integration times and number of measurements to be averaged to obtain precision of 0.1 % in $V(T)$ (see Supplementary Materials, Sec. 1).

As expected, the laboratory blackbody had the highest signal due to its close-to-unity emissivity. Except for sapphire when $T > 473$ K, all normalized voltages increase with temperature, even for the laboratory blackbody with a wavelength-independent emissivity, which is mainly due to the shape of $\eta(\lambda)$ (i.e., a finite detector bandwidth). Unlike the case in Fig. 1G, where the local slope of normalized power is determined by $\epsilon(\lambda)$, the shapes of the experimental normalized voltages are determined by $\epsilon(\lambda)\eta(\lambda)$.

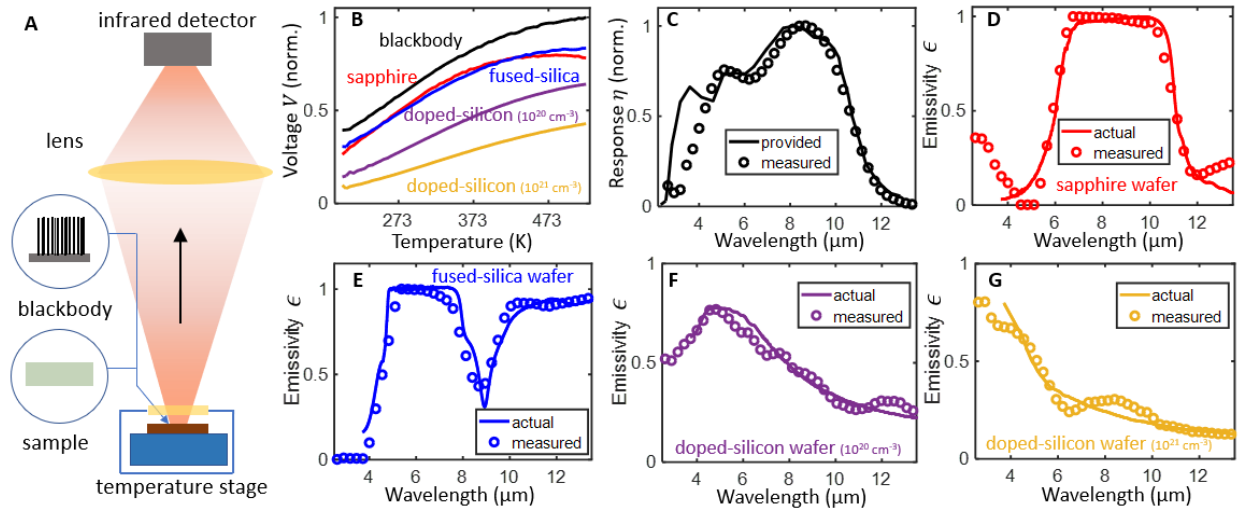

**Fig. 2. Experimental demonstration of Planck spectroscopy.** (A) Schematic of the experimental setup. Thermal emission from samples on a temperature stage was collected using a lens and focused onto an infrared detector. (B) Normalized voltage vs. temperature for five different samples: a laboratory blackbody, a sapphire wafer, a fused-silica wafer, and two n-doped silicon wafers with doping levels of $10^{20}$ and $10^{21}$ cm$^{-3}$. (C) The extracted response function of the measurement setup, obtained by using the blackbody reference (dotted circle), is in good agreement with the expected response function based on vendor-provided properties of the detector, heater window, and lens. (D-G) Measured emissivity of the samples using Planck spectroscopy (circles), and the corresponding spectra measured using a Fourier-transform spectrometer (solid line).

We used the laboratory blackbody reference (black curve in Fig. 2B), to extract the system response function $\eta(\lambda)$ (Fig. 2C), and the result agrees quite well with the expected spectrum, which is the product of the detector response (provided by the vendor), and the transmittance of the heater window and the lens. Using this calibrated $\eta(\lambda)$, we extracted the spectral emissivity for our four test samples, plotted in Figs. 2(D-G) together with their actual values measured using an FTS (see Supplementary Materials, Sec. 3). The emissivity values measured using Planck spectroscopy agree well with those made using an FTS and have an average mean-squared error (MSE) of about 0.007.

More generally, the accuracy of Planck spectroscopy depends on both the measurement precision of $V(T)$ and the range of measurement temperatures. In principle, a higher accuracy can be realized using a detector with higher responsivity and lower noise (e.g., liquid-nitrogen-cooled detectors [25]). For example, we estimate that the expected MSE can be reduced to 0.004 with an improved measurement precision in $V(T)$ of 0.01 % (see Supplementary Materials, Sec. 5).

The spectral resolution of Planck spectroscopy is not as straightforward to quantify as for grating or Fourier-transform spectrometers. Based on the experiments in Fig. 2, we achieved an approximate resolution of 1 µm. Using numerical simulations with measurement precision of 0.01 % and a temperature range of 173 to 523 K, Planck spectroscopy can capture an isolated peak or dip of about 0.4 µm in spectral width (see Supplementary Materials, Sec. 6). When two peaks are close to each other, Planck spectroscopy with the aforementioned measurement precision can resolve the two-peak feature if the peak separation is larger than 2 µm. We note that both the accuracy and the spectral resolution depends on the extraction algorithm, and better performance is likely achievable using more-sophisticated algorithms and additional constraints to the solution [6], [7], such as the use of an oscillator model like those used in ellipsometric analysis [26]. The accuracy and resolution may also be improved by the introduction of one or more optical filters into the setup, which may be viewed as a hybrid between Planck spectroscopy and filter-based spectroscopy

Though Planck spectroscopy is not directly applicable to samples whose emissivity changes significantly with temperature or that may be damaged at high temperatures, the simple setup in Fig. 1(D) can be modified slightly to avoid heating the sample, by placing a known reference sample on the heat stage and measuring light reflected or transmitted through the sample. The introduction of polarization elements into such a setup can also enable spectroscopic ellipsometry (see Supplementary Materials, Sec. 7).

In summary, we experimentally demonstrated and validated Planck spectroscopy—a spectroscopic technique that requires fewer optical components than any other existing approach. Planck spectroscopy does not require gratings, filters, or interferometers; instead, it uses the wavelength and temperature-dependent nature of the Planck blackbody distribution to acquire emissivity spectra of unknown samples. We envision implementations of Planck spectroscopy using infrared cameras to enable low-cost infrared hyperspectral imaging and imaging ellipsometry that does not sacrifice spatial resolution for spectral resolution.

**Funding:** We acknowledge support from the Office of Naval Research (N00014-20-1-2297).

**Author contributions**: Y.X. and M.K. conceived the project and designed the experiments. Y.X. carried out the experiments and performed the numerical calculations. C.W. and J.S contributed to setting up the experiment. I.M. contributed to the analysis of inversion of linear ill-conditioned equations. All authors discussed the results. Y.X. and M.K. wrote the manuscript with contributions from all other coauthors. M.K. supervised the project.

**Competing interests:** Authors declare no competing interests

**Data and materials availability:** All data is available in the main text or the supplementary materials.

**Supplementary Materials:**
Section 1: Measuring temperature-dependent thermal emission power
Section 2: Measuring sample surface temperature using an infrared camera
Section 3: Measuring sample emissivity using an FTS
Section 4: Obtaining spectral information from the integrated total emission power
Section 5: Extraction accuracy as a function of measurement accuracy and temperature range
Section 6: Spectral resolution
Section 7: Measuring temperature-dependent sample and enabling ellipsometry

# Supplementary Materials for

# Planck spectroscopy

Yuzhe Xiao[1], Chenghao Wan[1,2], Jad Salman[1], Ian J. Maywar[1], Jonathan King[1], Alireza Shahsafi[1], and Mikhail A. Kats[1,2,3*]

[1]Department of Electrical and Computer Engineering, University of Wisconsin-Madison, Madison, Wisconsin 53706, USA

[2]Department of Materials Science and Engineering, University of Wisconsin-Madison, Madison, Wisconsin 53706, USA

[3]Department of Physics, University of Wisconsin-Madison, Madison, Wisconsin, USA 53706

*Correspondence to: mkats@wisc.edu.

**This PDF file includes:**

Section 1: Measuring temperature-dependent thermal-emission power

Section 2: Measuring sample surface temperature using an infrared camera

Section 3: Measuring sample emissivity using an FTS

Section 4: Obtaining spectral information from the integrated total emission power

Section 5: Extraction accuracy as a function of measurement precision and temperature range

Section 6: Spectral resolution

Section 7: Measuring temperature-dependent samples and enabling ellipsometry

1. **Measuring temperature-dependent thermal-emission power**

The detector used in this study is a thermoelectrically cooled HgCdTe (MCT) detector from Boston Electronics (model: PVI-4TE-10.6) with a bandwidth from 3-11 μm and active area of $0.5 \times 0.5$ mm$^2$. The temperature stage is from Linkam Scientific (model: FTIR600), which has a temperature range from 78 to 873 K, and can be sealed using a barium fluoride (BaF$_2$) window. Samples were fixed onto the heater stage using kapton tape, with their thermal emission collected using a zinc selenide (ZnSe) lens (focal length of 25 mm) and focused onto the detector. The lens-to-sample and lens-to-detector distances were about 120 and 33 mm, respectively. This imaging system results in a measurement spot size of about $1.8 \times 1.8$ mm$^2$ on the sample, which is much smaller than the average sample size of $10 \times 10$ mm$^2$ used in this experiment.

In the experiment, thermal emission from samples with temperatures from 193 to 523 K were measured. The total thermal-emission signal decreases dramatically as the temperature is decreased, especially for temperatures below ambient. To obtain precise measurements, we used long integration times for the lower temperatures. However, there was non-negligible detector drift within the measurement time. To solve this problem, an optical chopper was placed in front of the sample with a low rotation speed of 0.2 rev/s. Then, the emission difference between the sample ("on" state, when the chopper blade did not block the sample) and the ambient-temperature chopper blade ("off" state, when the chopper blade blocked the sample) was used. The detector drift happened on a time scale of a few minutes or longer. Therefore, the measured voltage difference between adjacent "on" and "off" state (which is within about 5 seconds) is robust against the detector drift.

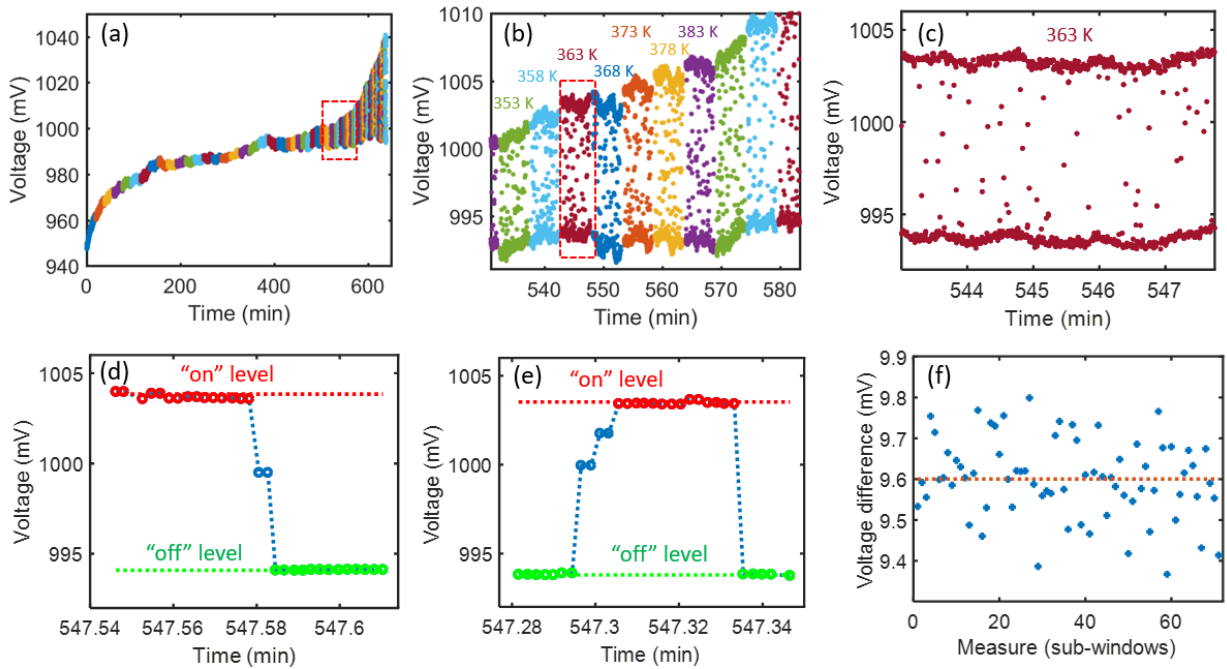

Figure S1: (a) Measured voltage for the laboratory blackbody reference. Data with different colors corresponds to measurements at different sample temperatures, ranging from 193 to 523 K, in steps of 5 K. The drift of the detector can be clearly seen from this measurement. (b) Zoom-in of the data in (a), showing the measurement for temperatures from 353 to 393 K. (c) Data for sample temperature of 363 K. The higher voltages correspond to the "on" state (the sample was fully unblocked), while the lower voltages correspond to the "off" state (chopper blade blocking the sample). Data points scattered

between these two regions correspond to the case where the chopper blade partially blocks the sample. (d-e) show how the data was analyzed in sub-windows. Each sub-window contains about 30 data points. The "on" and "off" levels were first determined by taking the average of five highest and lowest data points. Then data associated with "on" and "off" states (red and green circles) were decided by selecting data points that are within ± 0.5 mV of the "on" and "off" levels. Finally, the voltage difference was obtained using the mean value of the "on" and "off" states data points for each sub-window. (f) Voltage difference measured in 70 sub-windows for the data shown in (c), with the mean value shown by the red dotted line.

Figure S1(a) shows one such measurement for the laboratory blackbody reference, where the detector drift is obvious. The sample was first cooled to 193 K and then maintained for 15 mins to ensure the sample was fully cooled down. After that, the measurement began. Data points with different colors correspond to different sample temperatures. A non-uniform integration time was used since the thermal-emission power depends strongly on the sample temperature. The following measurement times were used: 20 mins per temperature for 193 to 273 K and 303 to 323 K, 10 mins per temperature for 328 to 373 K, 5 mins per temperature for 378 to 423 K, 3 mins per temperature for 428 to 473 K, and 2 mins per temperature for 478 to 523 K. The temperature window of 278 to 298 K was skipped because the signal (i.e., the difference of thermal emission between the sample and the ambient-temperature chopper blade) was particularly small in this temperature range. Figure S1(b) shows the zoomed-in portion of the measurement for sample temperature near 373 K, and (c) shows the data for sample temperature of 363 K. The data points on the top of (c) correspond to the signal when emission from the sample was measured ("on" state), while those on the bottom correspond to the signal when the sample was blocked by the chopper blade ("off" state). Data points scattered between these two regions correspond to the case where the chopper blade partially blocked the sample.

To obtain the difference in thermal emission from the chopper blade and the sample from these measurements, we break the measurement into small sub-windows, each containing about 30 data points (roughly one "on" state and one "off" state), as shown in Fig. S1(d-e). Within each sub-window, we first find the "on" and "off" levels by taking the average of the five highest and lowest data points. We then identify measurements that are in "on" and "off" states by picking data points that are within ± 0.5 mV (typical detector fluctuations are rarely larger than this value) of the "on" and "off" levels, which are shown by the red and green circles in (d-e), respectively. Finally, the voltage difference between the "on" and "off" states is obtained by taking the difference of the mean value for the "on"- and "off"-state data points. Figure S1(f) plots the voltage difference measured for 70 sub-windows from the data shown in (c). The mean value (red-dotted line) was then chosen to be the measured voltage difference for sample temperature of 363 K.

Figure S2 shows the measured voltage differences for (a) the laboratory blackbody reference, (b) the sapphire wafer, (c) the fused-silica wafer, and (d, e) two doped-silicon wafers, with doping level of (d) $10^{20}$ cm$^{-3}$ and (e) $10^{21}$ cm$^{-3}$. The doped-silicon wafer at $10^{21}$ cm$^{-3}$ has the lowest emissivity and therefore the lowest thermal-emission signal. Therefore, for this sample, we performed the full measurement from low temperature to high temperature using the same measurement settings in Fig. S1(a) for 10 times and then took the average of these 10 measurements. For the doped-silicon wafer at $10^{20}$ cm$^{-3}$, we performed 8 full measurements and then took the average. For the other samples, we performed 4 full measurements and then took the average. The deviations of these measurements from the averaged values are plotted in (f-j). For samples with relatively high emissivity (a-c), the standard deviation of each measurement is about 0.25 %. When taking the average of these measurement, the standard deviation is about 0.1 %. For the two doped-silicon wafers, the standard deviation of each measurement was larger, but the measurement times were also increased to ensure the measurement precision was similar to that of the cases in (a-c).

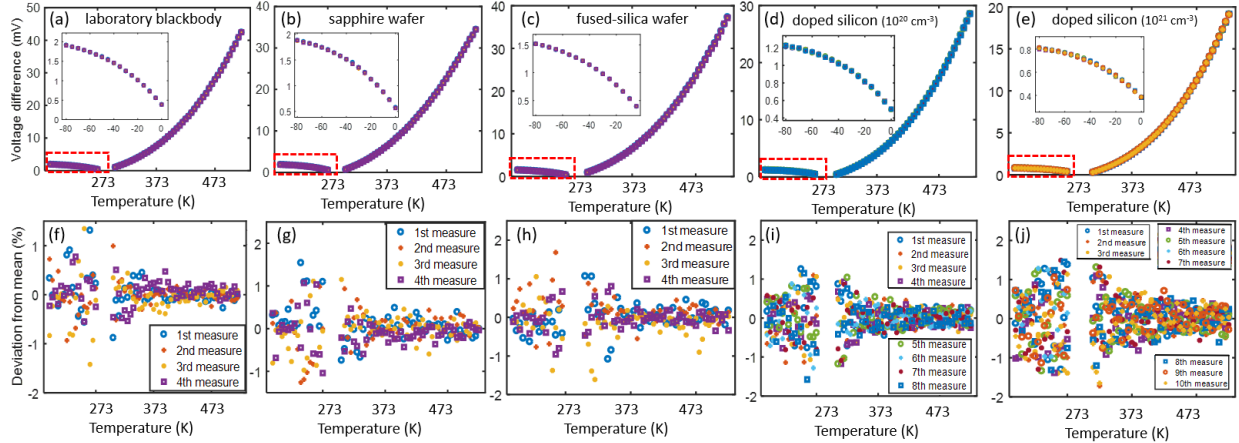

Figure S2: Measured voltage difference for (a) the laboratory blackbody reference, (b) the sapphire wafer, (c) the fused-silica wafer, and (d, e) two doped-silicon wafers, with doping level of (d) $10^{20}$ cm$^{-3}$ and (e) $10^{21}$ cm$^{-3}$. Due to the relatively lower signal, the measurements for doped-silicon wafer were repeated 8 and 10 times and then the averaged value of these measurements was taken to extract emissivity. The other samples measurements were repeated four times. (f-j) The measurement deviations from the mean value.

To better visualize the data, the experimentally measured voltage difference in Fig. S2 is normalized as:

$$V_{norm}(T) = \frac{V(T)-V(T_1)}{\alpha(T^4-T_1^4)} \tag{S1}$$

where $T_1 = 193$ K and the normalization factor $\alpha$ is chosen such that the normalized voltage of the laboratory blackbody reference at 523 K equals to one. The normalized voltages for all samples are plotted in Fig. 2B in the main text.

2. **Measuring sample surface temperature using an infrared camera**

Due to the finite thermal conductivity of the samples, there is a temperature difference between the heater stage and the top surface of the sample (*S1*). Such a temperature drop will lead to errors in the Planck-spectroscopy inversion process because inaccurate temperatures will be used in the blackbody matrix. We used a mid-infrared camera (FLIR A325sc, with a bandwidth from 7.5 to 13 μm) to measure the sample surface temperature. The infrared-camera software returns a map of temperature once a wavelength-integrated emissivity value is assigned in the camera software (we refer to this as $\epsilon_{set}$).

To measure the surface temperature, we first heated all samples to 323 K. At 323 K, the difference between the heater and ambient temperature is less than 30 K, resulting a relatively small temperature gradient between the top and bottom part of these samples, given the thickness and thermal conductivities of these samples (*S2*). So we adjusted $\epsilon_{set}$ such that the camera reading returned 323 K. We found $\epsilon_{set}$ of 0.97, 0.71, 0.86, 0.46, and 0.26 for the laboratory blackbody reference, the sapphire wafer, the fused-silica wafer, and the two doped-silicon wafers, respectively (Fig. S3, first column from left). Then the samples were further heated by setting the heater temperature to 373, 423, 473, and 523 K. The corresponding temperature readings from the camera are shown in Fig. S3. Due to the high thermal conductivity, there is almost no difference between the heater temperature and the surface temperature for the sapphire wafer and the doped-silicon wafer with doping level of $10^{21}$ cm$^{-3}$. For the laboratory blackbody and fused-silica wafer, their temperature drops are quite similar, with a value that is very close to our previous measurement (*S1*). For the doped-silicon wafer with doping level of $10^{20}$ cm$^{-3}$, a temperature drop of about 6 K was measured when it was heated to 523 K. We assume that this temperature drop mainly came from the contact resistance between the sample and heater because our doped silicon wafers are single-side polished with the unpolished side contacting the heater surface.

In the semitransparent region of a sample, the measured thermal-emission power not only comes from the top surface, but also has contributions from components beneath the surface. In an earlier work, we demonstrated such an effect by measuring the thermal-emission spectrum from a fused-silica wafer with a temperature gradient (*S1*). We want to note that such an effect is trivially small in the total emission power and not relevant for the experiment in this work.

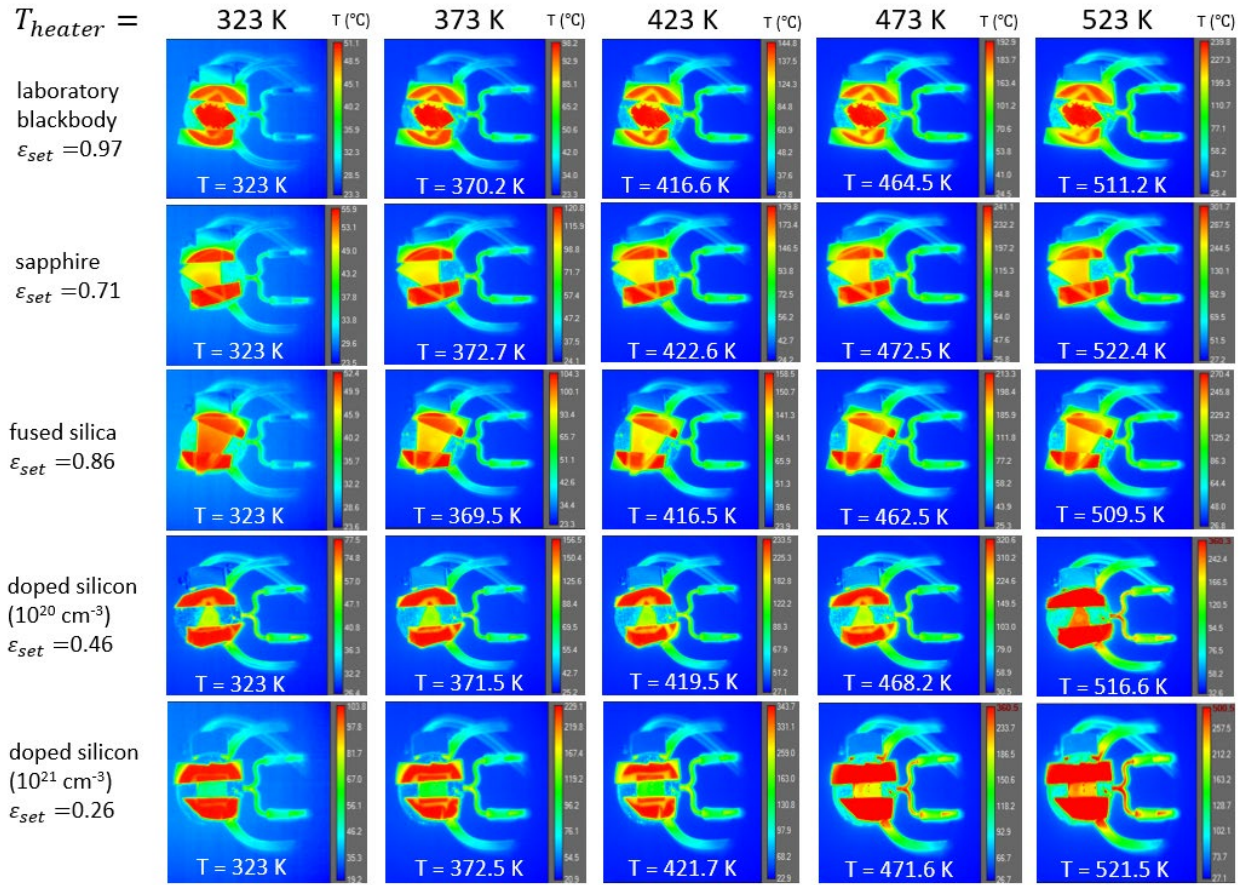

Figure S3: Measuring sample surface temperature with an infrared camera. The samples were first heated to 323 K, where $\epsilon_{set}$ of the camera was adjusted such that the camera read 323 K (first column from left). Then samples were heated to 373, 423, 473 and 523 K, where the readings from the infrared camera were measured to be the corresponding surface temperature.

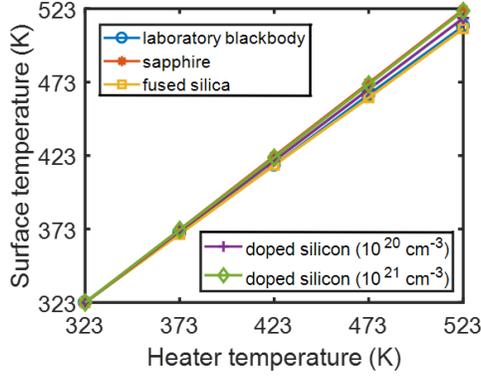

Figure S4: Measured surface temperature as a function of heater temperature for five different samples. The surface temperature follows a roughly linear relation with the heater temperature, as expected.

In Fig. S4, we plotted the measured surface temperature as a function of heater temperature for all five samples. The surface temperature roughly follows a linear relation with the heater temperature, which is expected. In finding the system response and the spectral emissivity, a linear temperature model was assumed for all samples:

$$T_{sample} = T_{heater} + \beta(T_{heater} - T_{room}), \qquad (S2)$$

where the slope coefficient $\beta$ is obtained from Fig. S4.

### 3. Measuring sample emissivity using an FTS

To validate Planck spectroscopy, we also measured the sample emissivity using a Fourier transform spectrometer (FTS, from Bruker, model: Vertex V70). Thermal emissivity of the sapphire and fused-silica wafers were measured in a previous study, where details can be found in ref. (*S2*). Here we show how we measured the emissivity of the two doped-silicon wafers.

The doped-silicon wafers were heated to two different temperatures and their emissivity was obtained using the following equation (*S2*):

$$\epsilon_x(\lambda) = \epsilon_{ref}(\lambda) \frac{S_x(\lambda, T_1) - S_x(\lambda, T_2)}{S_{ref}(\lambda, T_1) - S_{ref}(\lambda, T_2)}. \qquad (S3)$$

Here $S_x(\lambda, T)$ is the measured signal for sample $x$ at temperature $T$. We used the laboratory blackbody as the reference. Figure S5 (a-c) show the measured signal for these three samples at 323 and 353 K. The emissivity profiles for the two doped-silicon wafers calculated via Eq. S3 using the measurement in Fig. S5 are plotted in Fig. 2 in the main text.

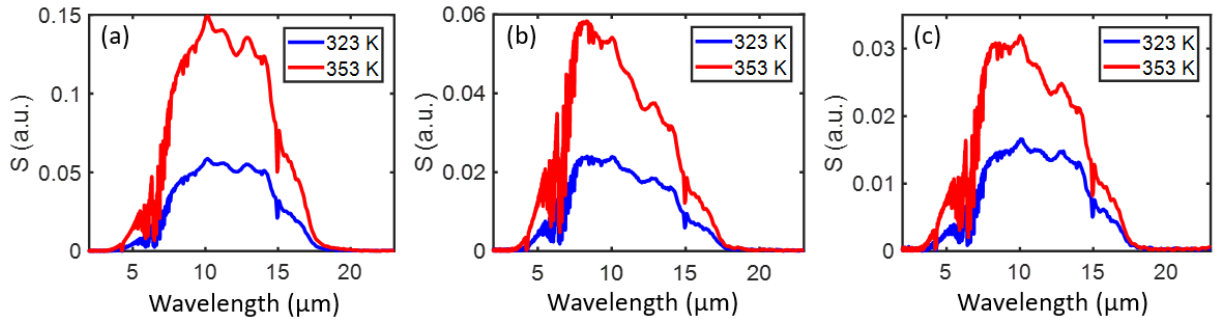

Figure S5: Measured thermal-emission signal using an FTS for the laboratory blackbody (a), doped-silicon wafer, $10^{21}$ cm$^{-3}$ (b), and doped-silicon wafer, $10^{21}$ cm$^{-3}$ (c).

## 4. Obtaining spectral information from the integrated total emission power

As discussed in the main text, using matrix inversion to solve Eq. 3 is not preferred because the blackbody-radiation matrix is ill-conditioned, and the solution is not robust against the noise in the measured voltage. For a linear matrix equation as follows:

$$\boldsymbol{y} = \boldsymbol{B}\boldsymbol{x}, \tag{S4}$$

the fluctuations $\boldsymbol{\delta x}$ in the solution are proportional to $\boldsymbol{\delta y}$ through the condition number of the matrix $C(\boldsymbol{B})$ ( *S3*):

$$\boldsymbol{\delta x} \sim C(\boldsymbol{B})\boldsymbol{\delta y}. \tag{S5}$$

If the condition number of a matrix is much larger than one, very small fluctuations in $\boldsymbol{y}$ will lead to large fluctuations in the solution $\boldsymbol{x}$. For temperature and wavelength settings similar to our experiment, the condition number of the blackbody-radiation matrix in Eq. 3 in the main text is on the order of $10^{10}$, making this equation very sensitive to the noise in the measured voltage.

Fortunately, there are constraints we can apply to Eq. 3 to achieve a more-robust solution: (i) $\epsilon(\lambda)$ is bounded between 0 and 1, and (ii) $\eta(\lambda)$ must be larger than 0. We used a linear least-squares solver with these bounds (the "lsqlin" solver in Matlab$^{TM}$). For this particular solver, we found that using a random portion of the data to solve Eq. 3 and then taking the average value of the solutions from these different random data portions returns a more robust solution than just solving all the data points at the same time. Therefore, we divided the experimental temperature points into 10 sub-windows of 30 K, with each sub-window containing 6 data points. Then, we picked 5 data points randomly from each of the 10 sub-windows (a total of 50 data points selected semi-randomly from a total of 60 data points) to solve for the spectral emissivity or response at 40 wavelength points. We performed calculations for 100 different random selections and took the average of these 100 different solutions to obtain the final solution.

The total thermally emitted power increases with temperature. Therefore, solving Eq. 3 directly will put more weight on the high-temperature measurements (because $\Delta V(T_j)$ is larger), effectively adding more weights in the short-wavelength region due to the blue shift of $I_{BB}(\lambda, T)$ for higher temperatures. Therefore, we solved the normalized version of Eq. 3 in the main text, taking the following form:

$$\begin{bmatrix} 1 \\ \vdots \\ 1 \end{bmatrix} = \begin{bmatrix} \Delta I_{BB}(\lambda_1, T_1)/\Delta V(T_1) & \cdots & \Delta I_{BB}(\lambda_N, T_1)/\Delta V(T_1) \\ \vdots & \ddots & \vdots \\ \Delta I_{BB}(\lambda_1, T_N)/\Delta V(T_N) & \cdots & \Delta I_{BB}(\lambda_N, T_N)/\Delta V(T_N) \end{bmatrix} \begin{bmatrix} \epsilon(\lambda_1) \\ \vdots \\ \epsilon(\lambda_N) \end{bmatrix} \begin{bmatrix} \eta(\lambda_1) \\ \vdots \\ \eta(\lambda_N) \end{bmatrix} \Delta\lambda. \tag{S6}$$

Finally, we smoothed the solution by averaging over the nearest 5 wavelength points (about 1 μm window), because we generally expect emissivity and response profiles to be smooth functions of wavelength. The smoothing performs a similar role to regularization. Also, as discussed in Sec. 6 of Supplementary Materials, the resolution of Planck spectroscopy is on the order of 1 μm, and therefore adding such smoothing does not sacrifice the resolution.

The steps of the inversion process are shown in Fig. S6. We plot the extracted system response from 100 random selections of the laboratory blackbody reference data in (a), and the correspondingly calculated normalized voltages from these solutions in (f). In (b-e), we plot the emissivity profiles extracted using 100 random selections of experimental data for the sapphire wafer, the fused-silica wafer, and the two doped-silicon wafers, while the correspondingly calculated normalized voltages are shown in (g-j), respectively. As shown in (a-e), the 100 individual solutions show some difference from each other, but the average values from these individual solutions (black lines) agree well with the actual values (Fig. 2 in the main text).

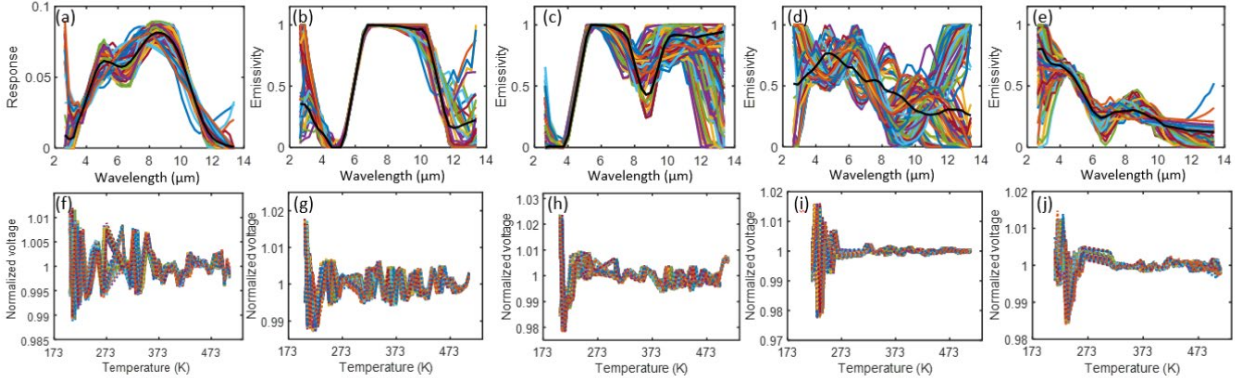

Figure S6: Extracted spectral emissivity from 100 random combinations of the experimental data, with the averaged value shown by the black curves, for the laboratory blackbody reference (a), sapphire (b), fused silica (c), doped-silicon with doping level $10^{20}$ (d) and $10^{21}$ cm$^{-3}$ (e). Bottom figures (f-j) plot the correspondingly fitted normalized voltage using the solutions shown on the top.

5. **Extraction accuracy as a function of measurement precision and temperature range**

The accuracy of Planck spectroscopy depends on a number of factors. The first factor is the measurement precision. The spectral emissivity is obtained by inversion of an ill-conditioned equation, which is unstable against noise in the measurement power. We performed numerical simulations to evaluate the robustness of Planck spectroscopy against measurement noise.

We chose a spectral bandwidth from 3 to 13 μm, which is the same as our experiment in the main text. We assumed a detector response that has a gaussian shape centered at 8 μm, with a bandwidth of about 5 μm, as shown in Fig. S7(a). Since the solver may perform differently for different emissivity profiles, we performed simulations for different randomized spectral emissivity profiles. More specifically, we use the following model to represent various sample emissivities:

$$\epsilon(\lambda) = \sum_{i=1}^{i=4} A_i e^{-(\lambda-\lambda_i)^2/\Delta\lambda_i^2}, \quad (S7)$$

where $A_i$, $\lambda_i$, and $\Delta\lambda_i$ are the weights, central wavelength, and bandwidth of each oscillator. These oscillator parameters were randomly selected within a specific range according to Table 1. The test spectral emissivity profile using the model in Eq. S7 with the choice of parameters in Table 1 reasonably resemble mid-infrared emissivity profiles from typical objects.

| Oscillator | 1 | 2 | 3 | 4 |
|---|---|---|---|---|
| $\lambda_i$ (μm) | [1 2] | [3 6] | [6 10] | [10 13] |
| $\Delta\lambda_i$ (μm) | [3 8] | [2 6] | [2 6] | [3 10] |
| $A_i$ | [0 1] | [0 1] | [0 1] | [0 1] |

Table S1: Range of parameters for the emissivity model that is used to evaluate the robustness of Planck spectroscopy against measurement noise.

Since the value of any spectral emissivity cannot exceed one, the emissivity in Eq. S7 is then normalized such that its peak value is between 0.2 and 1. For each emissivity profile, the voltage was calculated using Eq. 2 in the main text (assuming no background, which we expect to cancel out in the experiment by taking the voltage difference) for the temperature range from 193 to 523 K with a step size of 5 K. To mimic experimental measurement error, random noise with different relative magnitude $|\gamma|$ was then added to the exactly calculated signal:

$$V_{noisy}(T) = V_{exact}(T)(1 + \gamma). \tag{S8}$$

The calculated noisy voltage $V_{noisy}(T)$ was then used to solve for the spectral emissivity following the procedure discussed in Section 4. Simulations were performed for different magnitudes of noise, and 100 different random simulations were performed for each given noise level. Then, the mean-square error (MSE) of the extracted emissivity from these simulations were analyzed.

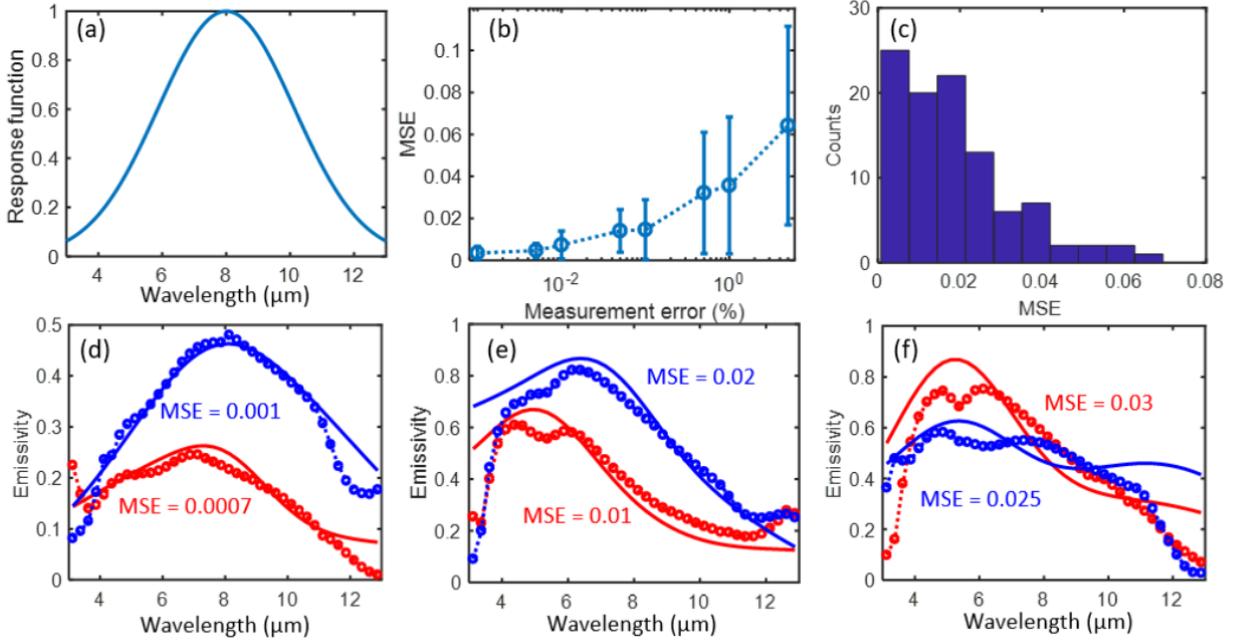

Figure S7: (a) The response function $\eta$ that is used to evaluate the robustness of the inversion process used in our implementation of Planck spectroscopy. (b) Averaged value and the standard deviation of the mean-squared error (MSE) of the spectral emissivity for different measurement error. (c) Histogram of the MSE for the 100 random simulations with a measurement noise level of $|\gamma| = 0.1\%$. (d-f) Several representative emissivity extractions in (c), with the input and extracted spectral emissivity shown by the solid and dotted lines, respectively.

The simulated averaged values of MSE and the standard deviation for different levels of measurement noise are plotted in Fig. S7(b). The MSE decreases quickly when the power-measurement precision is increased. Figure S7(c) shows the histogram of the MSE for 100 random simulations with a noise magnitude of $|\gamma| = 0.1\%$. For this level of measurement noise, the mean value of MSE is 0.015, with a standard deviation of 0.014. The majority of the MSE is smaller than 0.02. The experimental MSE of 0.007 demonstrated in this work falls well within the simulated range for the measurement error of 0.1%. Figs. S7(d-f) show several representative cases of the inversion process with noise magnitude of $|\gamma| = 0.1\%$, with the lowest MSE in (d) and the highest MSE in (f). For MSE smaller than 0.02, the spectral emissivity extracted using Planck spectroscopy is very robust.

The second factor that affects the accuracy of Planck spectroscopy is the measurement temperature range. This could be understood by considering spectrum reconstruction using a combination of filters: the reconstruction will be better for a bigger contrast between filters. The spectral shape of $I_{BB}(\lambda, T)$ changes with temperature. Therefore, a larger range of measurement temperature will lead to a bigger contrast in the shape of $I_{BB}(\lambda, T)$, which will improve the accuracy of Planck spectroscopy. This is especially true for lower temperatures: for a given temperature difference, the change in $I_{BB}(\lambda, T)$ is larger at low temperatures. Mathematically speaking, the inclusion of many different temperature points makes the linear problem less ill-conditioned.

We performed numerical simulations to demonstrate this effect. We fixed the number of power measurements to be 60, the spectral bandwidth to be from 3 to 13 μm, the highest temperature to be 523 K, the measurement noise $|\gamma|$ to be 0.1% and 0.01%, but changed the lowest temperature of measurement. For each lowest temperature, we performed 100 random simulations using the randomized input emissivity profile using Eq. S7. The averaged MSE from 100 random simulations, and the standard deviation are plotted in Fig. S8(a). The MSE decreases when the lowest temperature is reduced.

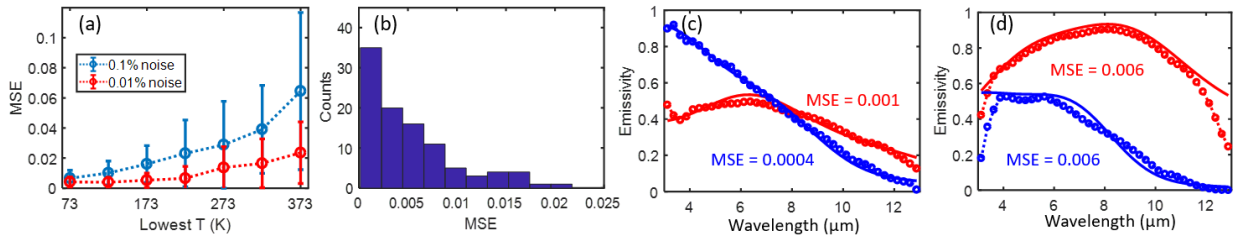

Figure S8: (a) Averaged values as well as the standard deviations of MSE for spectral-emissivity inversion for different value of lowest temperature with power-measurement noise level of 0.1% (blue) and 0.01% (red), respectively. (b) Histogram of the MSE for the 100 random simulations with measurement noise level of 0.01% and lowest measurement temperature of 173 K. (c-d) Several representative emissivity extractions in (b), with the input and extracted spectral emissivity shown by the solid and dotted lines, respectively.

For a measurement noise of 0.01% with the lowest measurement temperature of 173 K, the averaged MSE is 0.005, with a standard deviation of 0.0045. The corresponding histogram of MSE for 100 random simulations is shown in Fig. S8(b). Figs. S8(c-d) show several representative cases of spectral emissivity measurement from these simulations. As shown from Figs. S8(b-d), the performance of Planck spectroscopy can be very robust with realistic experimental conditions (0.01% measurement accuracy, which is feasible with a good infrared detector, and lowest measurement temperature of 173 K, which is feasible using a liquid-nitrogen cooling stage).

6. **Spectral resolution**

It is not easy to find materials with emissivity profiles featuring controllable narrow peaks in the mid infrared to test the resolution. Therefore, we tested the resolution of Planck spectroscopy with simulations. Here we assume the following realistic experimental conditions: measurement temperature from 173 to 523 K, with noise level of 0.01%. As in our experiments, we use the wavelength range of 3 to 13 μm and assume the detector response in Fig. S7(a).

We first test the ability of Planck spectroscopy to resolve a single narrow peak in the emissivity. We tested an emissivity profile with a single gaussian centered at 8 μm with different widths. In Fig. S9, we plotted the input and extracted emissivity profile. Based on these simulations, Planck spectroscopy can well measure a single peak about 0.4 µm wide. This is also partially confirmed by the experimental extractions of emissivity of the sapphire and fuse-silica wafers (Fig.2 in the main text), where sharp features such as the increase and decrease of sapphire emissivity near 6 and 11 µm, and the dip of emissivity of fused-silica near 9 µm, were well resolved.

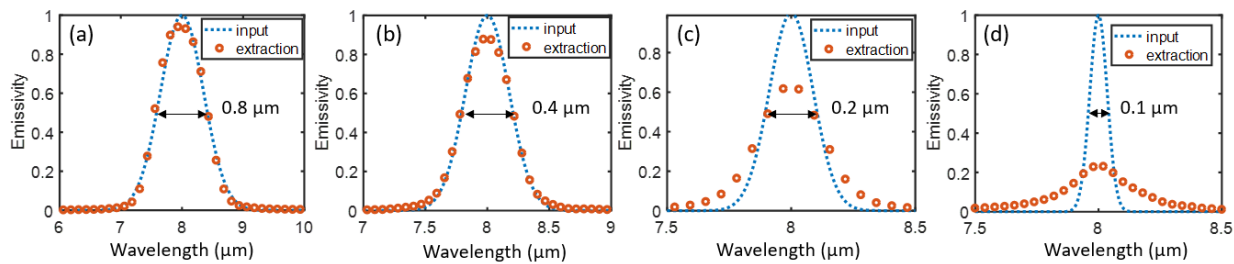

Figure S9: Measuring a single peak in emissivity using Planck spectroscopy. The widths of the gaussian peak (full width at half maximum) from (a) to (d) are 0.8, 0.4, 0.2, and 0.1 µm, respectively.

Another test of resolution is to resolve two nearby peaks. To test this, we considered an input emissivity that consists two gaussian peaks, each with full width at half maximum (FWHM) of 0.8 µm, and gradually decreased their separation from 3.5 to 2 µm. The results are plotted in Fig. S10. As shown in (a), when there are two nearby peaks in the emissivity profile, the extracted emissivity is not as good as the single-peak case (Fig. S9(a)). When the two-peak separation is 2.5 µm, we can barely resolve the two-peak feature. When the two-peak separation is further reduced to 2 µm, only one peak appears in the extracted spectrum.

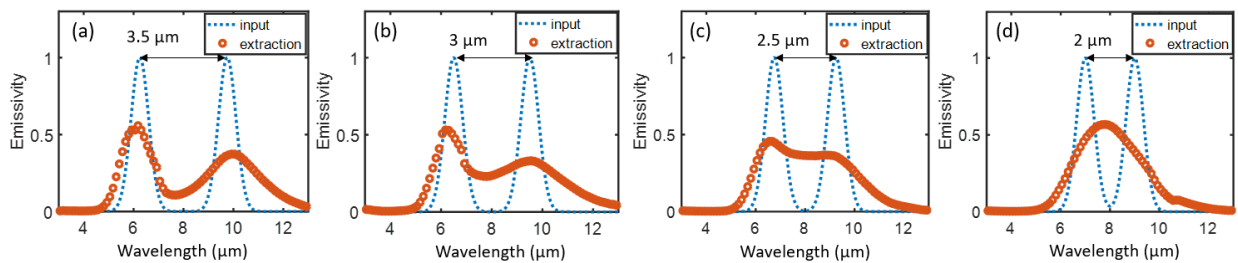

Figure S10: Resolving two closely spaced peaks in emissivity with Planck spectroscopy. The width of each of the input emissivity peaks (full width at half maximum) is 0.8 µm. The separations of two peaks in the input emissivity from (a) to (d) are 3.5, 3, 2.5, and 2 µm, respectively.

Both the accuracy and the spectral resolution depend on the measurement precision and the temperature range. Better performance can be expected if the measurement temperature range becomes larger. As an example, we performed similar simulations as those in Figs. S9-10, only reducing the lowest temperature to 73 K. Figure S11 shows the simulation results. In this case, Planck spectroscopy can well measure a single narrow peak down to a width of about 0.2 µm. Similarly, the cut-off separation for resolving the two-peak feature is reduced to 1 µm.

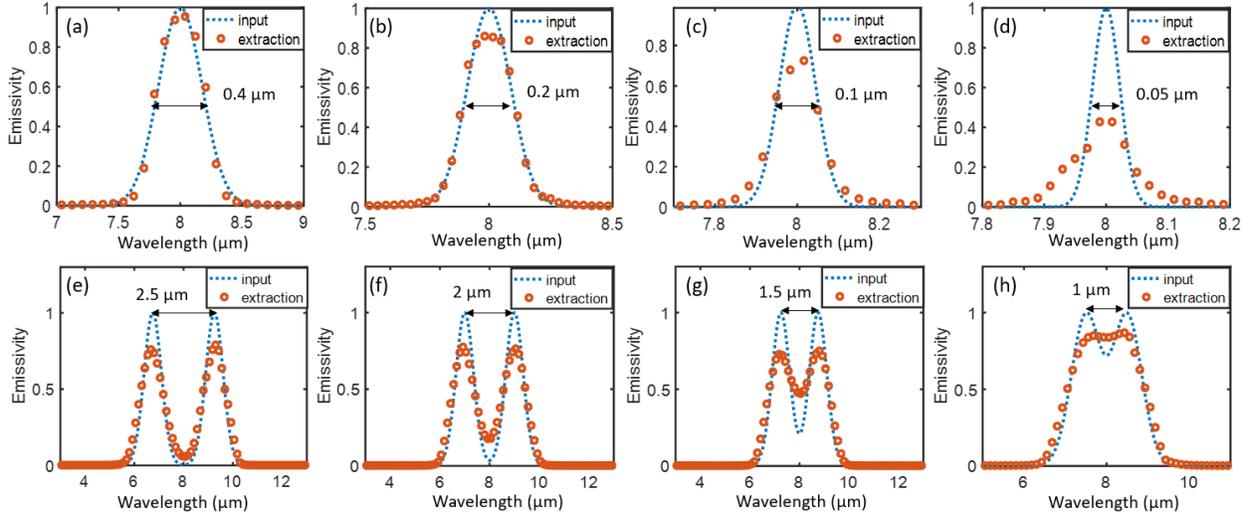

Figure S11. Top: measuring a single peak in emissivity with Planck spectroscopy. Same simulation condition with Fig. S9, except the lowest temperature is 73 K. The widths of the gaussian peak from (a) to (d) are 0.4, 0.2, 0.1, and 0.05 µm, respectively. Bottom: resolving two closely spaced peaks in emissivity. The simulations conditions are the same as in Fig. S10, except the lowest temperature is 73 K. The width of each of the input emissivity peaks is 0.8 µm. The separations of two peaks in the input emissivity from (e) to (h) are 2.5, 2, 1.5, and 1 µm, respectively.

7. **Measuring temperature-dependent samples and enabling ellipsometry**

The scheme of Planck spectroscopy described in Fig. 2 is not directly applicable to samples whose optical properties change significantly with temperature, or for fragile samples which cannot be heated at all. Figure S12 shows the schematic of a modified setup where one can use Planck spectroscopy to measure samples at a particular temperature.

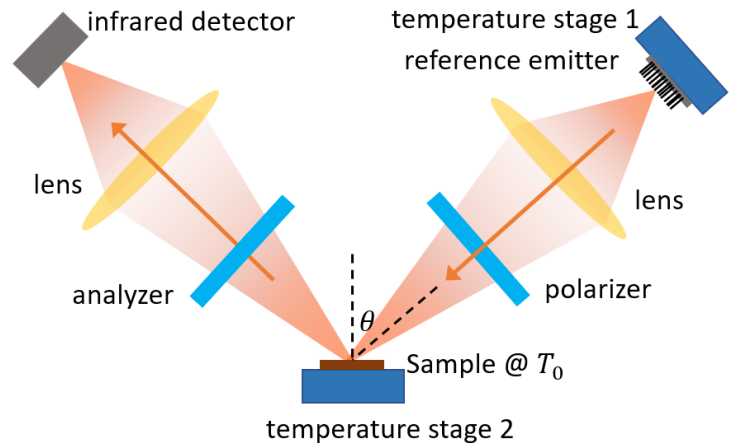

Figure S12. Schematic setup that can measure temperature-dependent samples with Planck spectroscopy. The inclusion of extra polarization elements enables spectroscopic ellipsometry.

Instead of measuring thermal emission from the sample, one can use the thermal emission from a known reference (such as a laboratory blackbody) and measure the reflectance or transmittance of the unknown sample. More specifically, in Fig. S12, temperature stage 1 changes the temperature of the reference emitter, which is needed for Planck spectroscopy, while temperature stage 2 controls the temperature of the sample to be characterized.

Additionally, extra polarization elements (e.g., a polarizer and an analyzer) can be included into the setup in Fig. S12, enabling the measurement of the sample reflection (or transmission) spectrum at different polarizations. This setup is the Planck-spectroscopy version of spectroscopic ellipsometry (*S4*).